\documentclass[aps,reprint,prl,twocolumn,floats,floatfix,longbibliography,superscriptaddress]{revtex4-1}

\usepackage{amsmath}
\usepackage{amsfonts}
\usepackage{amssymb}
\usepackage{graphicx}
\usepackage{color}
\usepackage[colorlinks=true,citecolor=blue,linkcolor=blue,urlcolor=blue]{hyperref}
\usepackage{times}
\usepackage{amstext}
\usepackage{latexsym}
\usepackage[running,mathlines]{lineno}
\usepackage{verbatim}
\usepackage[final]{pdfpages}
\usepackage{etoolbox} 

\makeatletter
\patchcmd{\@outputpage@head}{\@ifx{\LS@rot\@undefined}{}{\LS@rot}}{}{}{}
\makeatother

\providecommand{\moy}[1]{\langle #1 \rangle}

\providecommand{\ket}[1]{\lvert #1 \rangle}

\providecommand{\be}{\begin{equation}}
\providecommand{\ee}{\end{equation}}
\providecommand{\ba}{\begin{eqnarray}}
\providecommand{\ea}{\end{eqnarray}}

\begin{document}
	\raggedbottom

	\title{Multiphoton Jaynes-Cummings Model: Arbitrary Rotations in Fock Space and Quantum Filters}
	
	\author{Celso J. Villas-Boas}
	\email{celsovb@df.ufscar.br}
	\affiliation{Departamento de F\'{i}sica, Universidade Federal de S\~{a}o Carlos, 13565-905 S\~{a}o Carlos, S\~{a}o Paulo, Brazil}
	
	\author{Daniel Z. Rossatto}
	\email{dz.rossatto@unesp.br}
	\affiliation{Departamento de F\'{i}sica, Universidade Federal de S\~{a}o Carlos, 13565-905 S\~{a}o Carlos, S\~{a}o Paulo, Brazil}
	\affiliation{Universidade Estadual Paulista (Unesp), Campus Experimental de Itapeva, 18409-010 Itapeva, S\~{a}o Paulo, Brazil}

	\begin{abstract}
		
		The multiphoton Jaynes-Cummings model is investigated and applications in quantum information science are explored. Considering the strong atom-field coupling regime and an $N$-photon interaction, a nonlinear driving field can perform an arbitrary rotation in the Fock space of a bosonic mode involving the vacuum and an $M$-Fock state, with $M<N$. In addition, driving a bosonic mode with a linear coherent field (superposition of many Fock states), only the cavity states within the Fock subspace \{$\ket{0},\ket{1},\dots, \ket{N-1}$\} can be populated; i.e., we show how to implement a Fock state filter, or quantum scissor, that restricts the dynamics of a given bosonic mode to a limited  Hilbert space. Such a device can be employed as a generator of finite-dimensional quantum-optical states and also as a quantum-optical intensity limiter, allowing as a special case the generation of single-photon pulses. On the other hand, our system also provides a very rich physics in the weak atom-field coupling regime, in particular multiphoton electromagnetically induced transparencylike phenomena, inducing a narrow (controllable) reflectivity window for nonlinear probe fields. These results are useful for applications in quantum information processing and also motivate further investigations, e.g., the use of an $N$-photon Jaynes-Cummings system as a qudit with harmonic spectrum and the exploration of multiphoton quantum interference.
	\end{abstract}

	\maketitle
	\emph{Introduction.}---Recent technological advances in manipulating the radiation-matter interaction, especially at the level of a few atoms and photons, have allowed great strides in the implementation of quantum information processing protocols.  The mastery of such interactions on various platforms \cite{Haroche2013,Reiserer2015,Wineland2013,Hennessy2007,Devoret2013} has made it possible to dispel mistrust regarding the possibility of a quantum computer becoming real \cite{Haroche1996}. Thus, one often sees new (or improved) quantum computing protocols being implemented on diverse setups \cite{Palmero2017,Nguyen2018,Welte2018,Zajac2017,Axline2018,Boixo2018,wendin2017}. 
	
	The most elementary interaction between atom and radiation was first described by Rabi in 1936 \cite{Rabi1936}, considering the radiation as a classical field and a dipolar interaction. 
	In its quantum version, it is possible to classify such a model into different regimes \cite{Rossatto2017,FDiaz2018,Kocku2018}. When the interaction energy is a small perturbation on the free energies, the quantum Rabi model can be reduced to the well-known Jaynes-Cummings (JC) model \cite{Jaynes1963} through the rotating-wave approximation, which describes the usual coherent exchange of a quantum of energy between a two-level atom and a single-mode bosonic field.
	
	The JC model can be extended to nonlinear versions (nondipolar light-matter interaction) \cite{Shore1993}, in particular the multiphoton JC model \cite{Sukumar1981,Singh1982} that considers multiphoton exchange. Such a nonlinear Hamiltonian appears, at least for two-photon interaction, in trapped-ion domain \cite{Vogel1995,Meekhof1996}, optical \cite{Gau1992} and microwave \cite{brune1997} cavities, or even in superconducting circuits \cite{neil2015,Garziano2015,Bertet2005,Felicetti2018,Feli2018}. In Refs.~\cite{Feli2018,Felicetti2018}, a much more interesting scenario is presented since two-photon interactions can be implemented with substantial coupling strengths, differently from perturbative higher-order effects of a dipolar interaction that allow small effective coupling strengths \cite{Vogel1995,Meekhof1996,Garziano2015,Piazza2012,Hamsen2017}. Furthermore, new physical consequences emerge when the two-photon interaction becomes stronger and stronger \cite{Duan2016,Garbe2017,Chen2018}. The feasibility of achieving nonperturbative two-photon interactions and the possibility to broaden it for general multiphoton processes stimulate a pursuit of novel physical phenomena and applications to such nondipolar interactions, such as new ways to manipulate quantum information.

	In this Letter we investigate the multiphoton JC Hamiltonian ($N$-photon interaction) and how it can be employed in quantum information science. Given the system in the strong coupling regime, we show how to perform arbitrary rotations in Fock space involving vacuum and $M$-photon states ($M<N$), and how to implement a quantum device that allows the transmission of a field only in a finite superposition (or mixture) of Fock states (quantum scissor \cite{Leoski2011,Hoi2012,Peropadre2013}). In particular, the two-photon JC interaction can be used for single-photon generation in a different fashion as compared to protocols involving standard JC interactions \cite{Lindkvist2014,Peng2016}. Additionally, we show that the system can also provide a rich physics in the weak coupling regime, in particular multiphoton electromagnetically induced transparencylike phenomena; i.e., the multiphoton absorption of the bosonic mode can be canceled out due to a quantum destructive interference, inducing a narrow (controllable) reflectivity window for nonlinear probe fields.

\begin{figure}[b]
	\includegraphics[trim = 0mm 0mm 0mm 0mm, clip, width=0.48\textwidth]{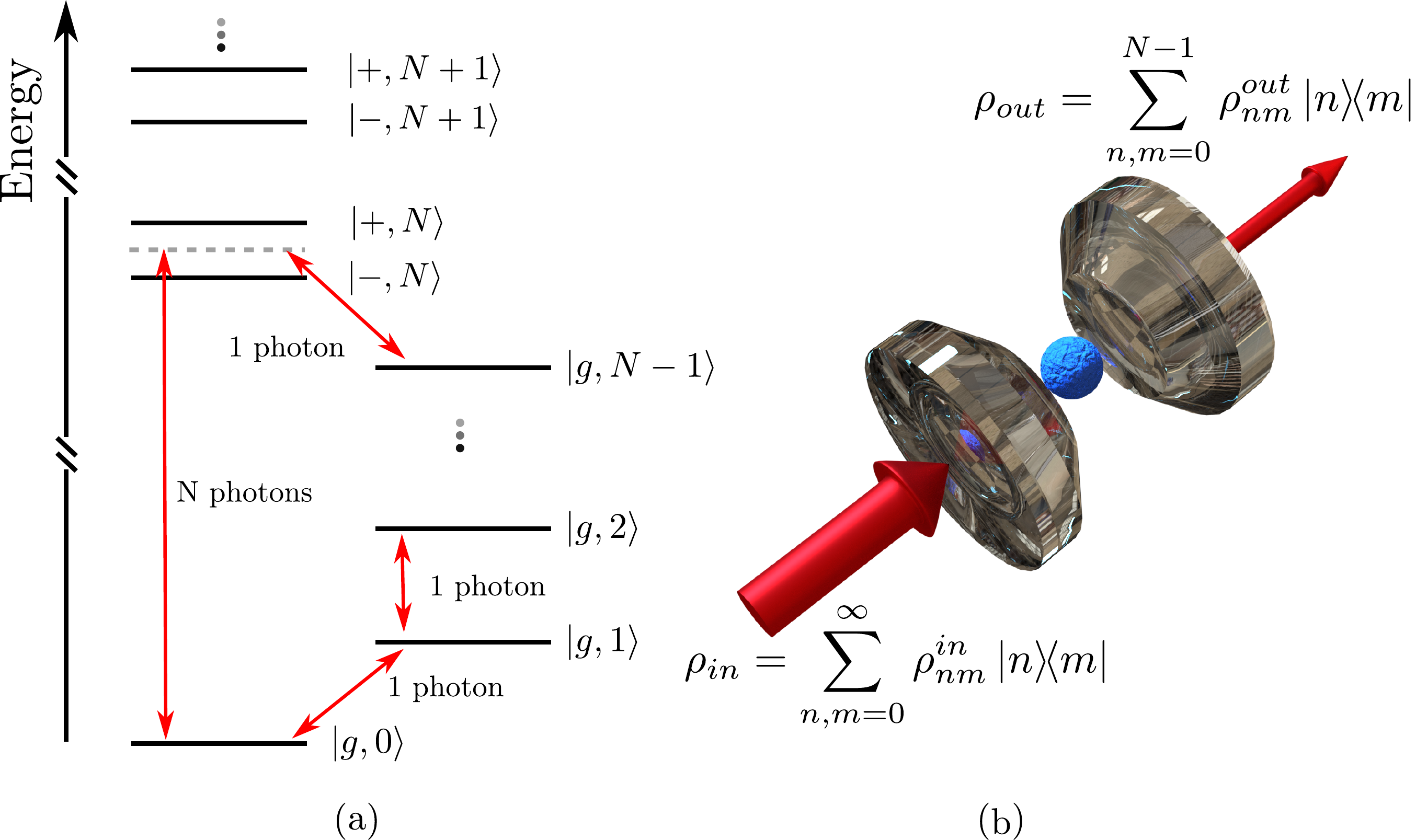} 
	\caption{(a) Energy spectrum of the $N$-photon Jaynes-Cummings model. The system has $N$ equidistant uncorrelated eigenstates, $\left|g,n\right\rangle $ for $0\le n \le  N-1$, and dressed eigenstates, $\left|\pm,n\right\rangle =\frac{1}{\sqrt{2}}\left(\left|g,n\right\rangle \pm\left|e,n-N\right\rangle \right)$ for $n \ge N$ (Jaynes-Cummings-like doublets). The transition $\ket{g,0} \to \ket{\pm,N}$ can be induced by an $N$-photon driving field on the cavity mode, while the adjacent uncorrelated states can be addressed via linear driving fields (single-photon transitions). (b) Pictorial representation of the driven atom-cavity system. Giving an input field, the output field will be limited to the ($N-1$)-photon state.}
	\label{fig1}
	\end{figure}

	\emph{Model.}---We consider the $N$-photon JC model \cite{Sukumar1981,Singh1982} ($\hbar=1$),
	\begin{equation}
	H_{0}=\omega a^{\dagger}a+\omega_{0}\frac{\sigma_{z}}{2}+g(\sigma_{+}a^{N}+ {\rm{H.c.}}),\label{hamiltonian}
	\end{equation}
	with $\omega$ and $a$ ($a^{\dagger}$) being the frequency and the annihilation (creation) operator of the single-mode bosonic field. The atomic frequency transition between the ground $\ket{g}$ and excited $\ket{e}$ states is $\omega_{0}$, while $\sigma_{+}=\left(\sigma_{-}\right)^{\dagger}=\left|e\right\rangle \left\langle g \right|$ and $\sigma_{z}=\sigma_{+}\sigma_{-}-\sigma_{-}\sigma_{+}$. The atom-field coupling is $g$ and ${\rm{H.c.}}$ stands for Hermitian conjugate. This Hamiltonian describes a coherent exchange of $N$ excitations of the bosonic mode with a two-level atom. In Fig.~\ref{fig1}(a) we illustrate the energy-level diagram for a resonant interaction ($\omega_{0}=N\omega$). The lowest eigenstates are uncorrelated, involving the atomic ground state and up to $N-1$ excitations in the bosonic mode, 
	\begin{equation}
	\left|\Psi_{g,n}\right\rangle =\left|g,n\right\rangle \,\,\text{for}\,\,\, 0 \le n < N,\label{groundstates}
	\end{equation}
	with eigenenergies $E_{g,n}=(n - \tfrac{N}{2} )\omega$. The remaining (dressed) eigenstates are 
	\begin{eqnarray}
	\left|\pm,n\right\rangle =\frac{1}{\sqrt{2}}\left(\left|g,n\right\rangle \pm\left|e,n-N\right\rangle \right) \,\,\text{for}\,\,\,  n \ge N,
	\label{excitedstates}
	\end{eqnarray}
	with eigenenergies $E_{\pm,n}=(n - \tfrac{N}{2} )\omega\pm g\sqrt{\tfrac{n!}{(n-N)!}}$. Therefore, the system exhibits a finite-dimensional harmonic-oscillator spectrum \cite{Miranowicz2001} for the lowest eigenstates and then followed by JC-like doublets, as depicted in Fig.~\ref{fig1}(a). For the sake of illustration, we hereafter consider the single-mode bosonic field as a mode of an optical cavity, as sketched in Fig.~\ref{fig1}(b).
	
	The dressed states can be directly excited from the ground state via a nonlinear driving field \cite{brune1997,neil2015,Andersen2016,Mallet2011} on the cavity mode. To this end, we assume a driving field of strength $\varepsilon(t)$, single-photon frequency $\omega_{p}$, phase $\chi$, and nonlinearity $M$, which is described by $H_{P}=\varepsilon(t)[a^{M}e^{i(M\omega_{p}t-\chi)}+{\rm{H.c.}}]$. Thus, the total Hamiltonian, written in a quasi-time-independent frame, reads
	\begin{equation}
	H_{I}=\Delta_{p}\left(a^{\dagger}a+N\frac{\sigma_{z}}{2}\right)+[\varepsilon(t)a^{M}e^{-i\chi}+ga^{N}\sigma_{+}+{\rm{H.c.}}],\label{eq:interactionhamiltonian}
	\end{equation}
	with $\Delta_{p}=\omega-\omega_{p}$. Finally, considering the cavity and the two-level atom coupled to their reservoir in the white-noise limit (Born, rotating-wave and Markov approximations in the system-reservoir interactions) \cite{gardiner2004,Petruccione2007}, the system dynamics at $T=0$K is governed by the master equation 
	\begin{equation}
	\begin{split}\frac{d\rho}{dt}=& -i[H_{I},\rho]+\gamma(2\sigma_{-}\rho\sigma_{+}-\sigma_{+}\sigma_{-}\rho-\rho\sigma_{+}\sigma_{-})\\
	& +\kappa(2a\rho a^{\dagger}-a^{\dagger}a\rho-\rho a^{\dagger}a)+\gamma_{\phi}(\sigma_{z}\rho\sigma_{z}-\rho),
	\end{split}
	\label{eq:masterequation}
	\end{equation}
	in which $\gamma$ and $\gamma_{\phi}$  are the atomic polarization decay and dephasing rates, respectively, while $\kappa$ is the decay rate of the cavity field amplitude. Moreover, this master equation is valid whenever we are out of the ultrastrong and deep-strong atom-cavity coupling regimes ($g\ll\omega,\omega_{0}$), and for small excitation numbers \cite{Blais2011}. We can numerically solve Eq.~\eqref{eq:masterequation} by truncating the Fock space of the cavity mode according to $\varepsilon$ and $M$ \cite{Johansson2012}.


	\emph{Strong coupling regime: Arbitrary rotation in Fock space and quantum scissor.}---First we analyze the system dynamics in the strong coupling regime; i.e., when the coherent atom-field interaction exceeds all relaxation processes ($g>\kappa,\gamma,\gamma_{\phi}$).
	Adjusting the driving field resonantly to the cavity mode frequency ($\omega_{p}=\omega$) and assuming the system in its ground state initially, two cases appear if the driving field is considered as a weak probe field [$\varepsilon_m \equiv \max{(| \varepsilon(t) |)} \ll  g$]. (i) When $M>N$, the dressed states of the system are not populated since the probe field is out of resonance with the transitions $\ket{g,0} \to \ket{\pm,M}$, also for $\ket{\pm,M} \to \ket{\pm,2M}$ and so on. Consequently, the system remains in the ground state while the driving field is completely reflected by the cavity mirror. (ii) When $M<N$,  a more interesting situation takes place, in which the probe field is able to induce the system to the uncorrelated states only. For instance, for $N=2$, the most excited uncorrelated state is $|g,1\rangle$, which can be populated with a linear probe field ($M=1$). Nevertheless, the two-photon state (and consequently the higher ones) is not populated since the transitions $|g,1\rangle \to |\pm,2\rangle$ are not resonant with this linear probe field, thus prohibiting the injection of more than one photon into the cavity mode (single-photon blockade \cite{[{For a recent review, see Section 6.6 of }][{, and references therein.}]Gu2017}). Hence, the system dynamics is restricted to the Fock subspace $\{|0\rangle,|1\rangle\}$, even if we take into account stronger driving fields, but ensuring that $g$ be strong accordingly. For different $N$ and $M$ ($<N$) similar situations appear, in which the system exhibits multiphoton blockade phenomena \cite{Gu2017}.

\begin{figure}[t]
		\includegraphics[trim = 6mm 2mm 1mm 1mm, clip, width=0.48\textwidth]{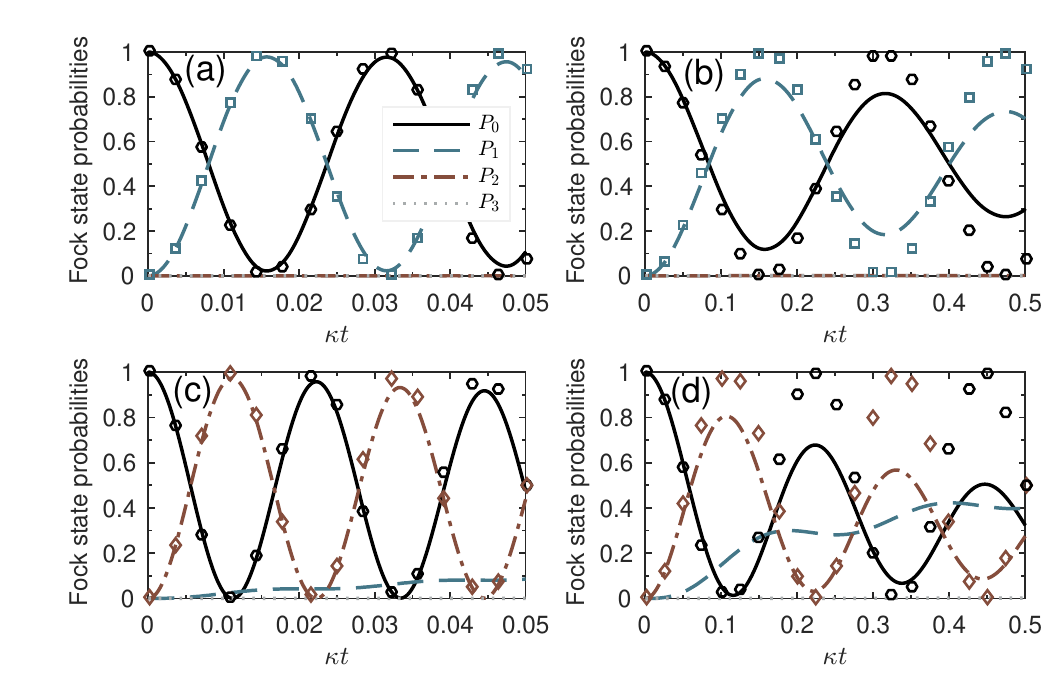}
		\caption{Rotations in Fock space involving the vacuum and $M$-Fock states. We plot the Fock state probabilities $P_{k}(t)={\rm{Tr}}[|k\rangle\langle k|\rho(t)]$ as a function of $\kappa t$ for $\gamma=\gamma_\phi = \kappa/2$ and $\chi=0$. In the upper panels $N=2$ and $M=1$, assuming (a) $g=1000\kappa$ and $\varepsilon(t)=\varepsilon_{0}=100\kappa$, and (b) $g=100\kappa$ and $\varepsilon_{0}=10\kappa$. We notice that the rotation fidelity can be high when $\varepsilon_{0} \gg \kappa$, but it decreases when $\varepsilon_{0} \gtrsim \kappa$ (cavity dissipation destroys the quantum superposition). In the lower panels  $N=3$ and $M=2$, with (c) and (d) given in terms of the parameters of (a) and (b), respectively. For this configuration, we can see that the cavity dissipation can also lead the cavity to a final state outside the desired Hilbert space \{$\ket{0},\ket{2}$\}, since the Fock state $\ket{1}$ is populated through the decay channel $\ket{2} \to \ket{1}$. The symbols correspond to the approximate solution [Eq.~\eqref{psi_rotating}].} 
		\label{Figure_rotations}
	\end{figure}

	Considering the last case, let us derive an effective Hamiltonian in the limit of atom-field coupling much stronger than the driving field strength, i.e., when $g \gg \varepsilon_m$ (weak probe field). To this end, we rewrite $H_{P}$ (for $M<N$) in the eigenbasis of $H_{0}$. Then, adjusting $\omega_{p}=\omega$ and performing a rotating-wave approximation \cite{SM}, we have 
	\begin{equation}
	H_{\rm{eff}} \simeq \varepsilon(t) e^{i\chi} \sum_{\substack{n=0 \\ (M<N)}}^{N-M-1}\sqrt{\frac{(n+M)!}{n !}}\left|g,n+M\right\rangle \langle g,n|+{\rm H.c.},\label{effective_hamiltonian}
	\end{equation}
	which is valid for $\varepsilon_m \ll g \sqrt{2\,(N-M)!}$ \cite{SM}. When $\frac{N}{2} \le M <N$, $H_{\rm{eff}}$ promotes an arbitrary rotation between the vacuum and the $M$-Fock states; i.e., $\ket{\Psi(t)} = e^{-i\int_{0}^{t}H_{\rm{eff}}(t^{\prime})dt^{\prime}}\ket{g,0}$ reads
	\begin{equation}
	\ket{\Psi(t)} =\ket{g} \left[\cos{\left(\frac{\theta_M(t)}{2}\right)\ket{0}}+e^{i\varphi}\sin{\left(\frac{\theta_M(t)}{2}\right)\ket{M}}\right],
	\label{psi_rotating}
	\end{equation}
	with $\theta_M(t)=2\sqrt{M!}\int_{0}^{t} \varepsilon(t^\prime)dt^\prime$ and $\varphi=\chi - \frac{\pi}{2}$ (polar and azimuthal angles of the Bloch sphere, respectively). As the atom remains in its ground state throughout the dynamics, this rotation gate is immune to atomic relaxation processes. On the other hand, the cavity dissipation introduces errors to the gate because it destroys the quantum superposition of Eq.~\eqref{psi_rotating} and also leads the cavity to a final state outside the desired Hilbert space \{$\ket{0},\ket{M}$\}. In this sense, the greater the ratio $\varepsilon_m/\kappa$ the higher the gate fidelity, since in this case the rotation tends to occur before the cavity dissipation appreciably disturbs the unitary evolution. Such detrimental influence on the rotation gate can be seen in Fig.~\ref{Figure_rotations}, in which we illustrate the time evolution of the Fock state probabilities for $N=2$ and $M=1$ (upper panels) and for $N=3$ and $M=2$ (lower panels).
	
	Therefore, we have shown a straightforward scheme to perform arbitrary rotations in Fock space (involving vacuum and $M$-Fock states), which could be implemented at least for $N=2$ with the current technology \cite{Feli2018,Felicetti2018}, moving toward a suitable route to unitary gates for Fock state qubits. We must remember that controlled qubit operations in Fock space are not a simple task \cite{Nielsen,Santos2005,Miled2014,Prado2014,Rosado2015,Krastanov2015,Heeres2015,Hofheinz2008,Liu2004,Law1996}, often demanding multistep processes and being sensitive to the dissipation of the atomic ancilla.

	A linear driving field is able to inject up to $N-1$ excitations into the cavity mode. For instance, for $N=3$, $M=1$, and $\varepsilon_m \ll 2g$, we have $H_{\rm eff}\simeq\varepsilon(t)e^{i\chi} \left|g\right\rangle \langle g|(\left|1\right\rangle \langle 0|+\sqrt{2} \left|2\right\rangle \langle 1|)+{\rm H.c.}$, which indicates that the cavity mode can only be populated in the vacuum, one-, and two-photon states. Thus, our system also works out as a quantum scissor, generating a finite-dimensional quantum-optical state inside the cavity, but without requiring a nonlinear Kerr medium inside it as usually adopted in the literature for this purpose \cite{Leoski2011}. Our case has the advantage that the multiphoton interaction can achieve strong strengths as shown in Refs.~\cite{Feli2018,Felicetti2018}.
	
	From the input-output theory, the field transmitted by a cavity mirror, opposite to the driven one, is exactly the intracavity field multiplied by the square root of the decay rate relative to this mirror \cite{walls2007quantum}. Then, from this perspective, our system also acts as a quantum-optical intensity limiter or Fock state filter; i.e., driving the cavity with an input field comprising any superposition (or mixture) of Fock states, the output field will be a finite superposition (or mixture) limited to the ($N-1$)-photon state, restricting the maximum intensity of the transmitted field [see Fig.~\ref{fig1}(b)]. If one wants to ensure that the field be almost entirely transmitted by a specific mirror, it is convenient to employ an asymmetric cavity \cite{Mcke2010}.
	
	\begin{figure}[t]
		\includegraphics[trim = 6mm 8mm 1mm 4mm, clip, width=0.48\textwidth]{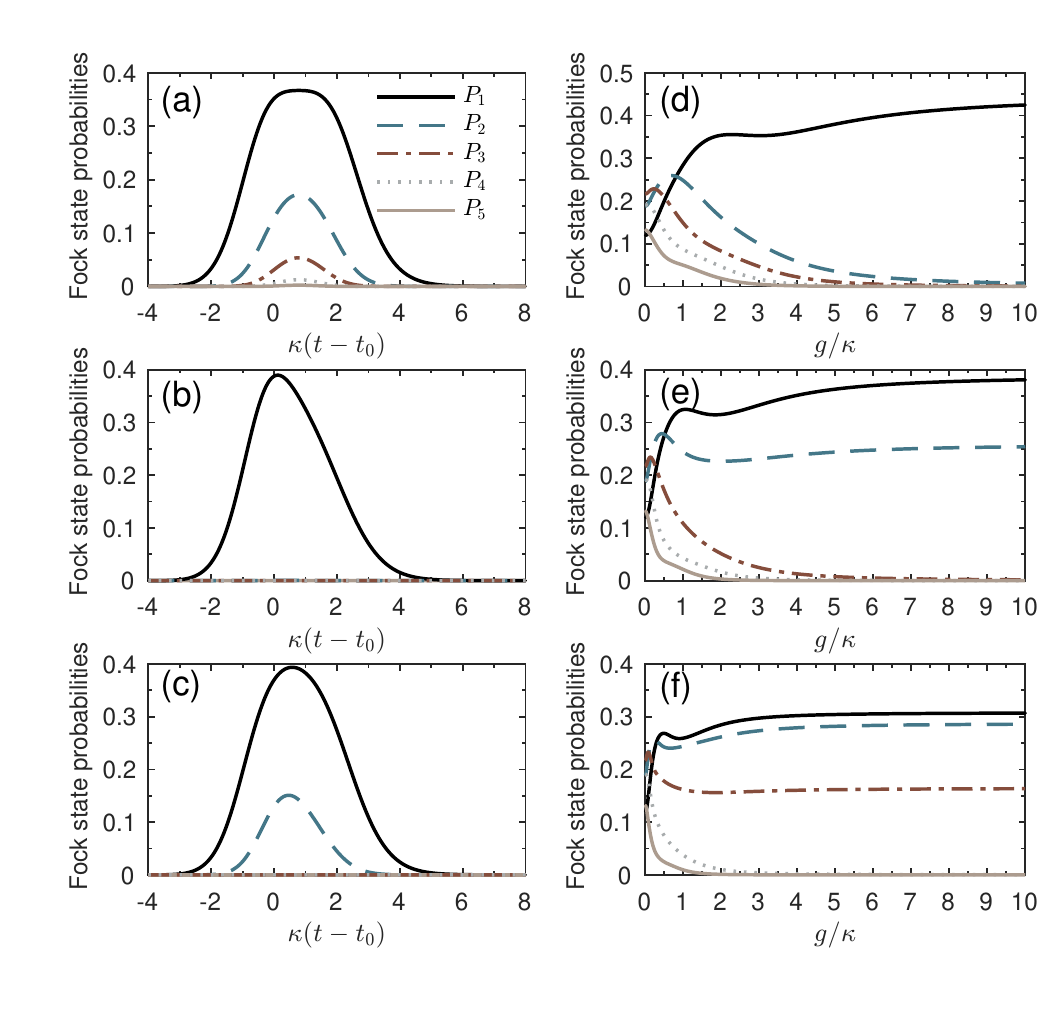}
		\caption{Fock state filter. In the left-hand panels we plot the Fock state probabilities $P_{k}(t)={\rm Tr}[|k\rangle\langle k|\rho(t)]$ as a function of $\kappa (t-t_0)$ assuming a Gaussian probe pulse ($M=1$, $\varepsilon_{m}=4\kappa$ and $\eta=\sqrt{2}\kappa^{-1}$) and $\gamma=\gamma_{\phi}=\kappa/2$ for (a) $g=0$ and $g=20\kappa$ [(b) $N=2$ and (c) $N = 3$]. In the right-hand panels we plot $P_{k}$ in the stationary regime as a function of $g/\kappa$ assuming a continuous driving field with $\varepsilon(t)=\varepsilon_{0}=2\kappa$ for (d) $N=2$, (e) $N=3$, and (f) $N=4$.}
		\label{Figure_filter}
	\end{figure}
	
	As mentioned above, the determination of the density matrix of the intracavity field is enough to derive the properties of the output field. Thus, let us first consider the driving field as a Gaussian pulse with amplitude $\varepsilon(t)=(\varepsilon_{m}/\sqrt{2\pi\eta^{2}})e^{-(t-t_{0})^{2}/2\eta^{2}}$, in which $\varepsilon_{m}$ and $\eta$ are the maximum amplitude and duration of the driving pulse, respectively, and $t_{0}$ is the time when its maximum arrives at the cavity mirror. For $g=0$ (driven empty cavity), we have non-null probabilities ($P_k$) of finding $k$ excitations in the cavity mode, as we can see in Fig.~\ref{Figure_filter}(a) considering $\varepsilon_{m}=4\kappa$ and $\eta=\sqrt{2}\kappa^{-1}$. However, depending on $N$, only some Fock states are populated when $g \neq 0$. For $N=2$ ($3$), the highest cavity-mode state populated is $|1\rangle$ ($|2 \rangle$) [Figs.~\ref{Figure_filter}(b) and \ref{Figure_filter}(c)], such that we can have a single-photon source for $N=2$.  For weak atom-field coupling and intense driving fields, this quantum filter does not work out since the driving field is able to introduce more excitations in the cavity mode, as observed in Figs.~\ref{Figure_filter}(d) and \ref{Figure_filter}(f) for $g \lesssim \kappa , \varepsilon_{m}[2(N-M)!]^{-1/2}$. These figures show $P_k$ as a function of $g/\kappa$ in the steady state assuming a continuous driving field. We observe that $P_{k \ge N} \to 0$ as the system reaches stronger couplings ($g \gtrsim 10\kappa$), which elucidates the truncation of the Fock space accessible to the cavity mode. It is worth remarking that the Fock state filter yields similar results for nonlinear driving fields ($1<M<N$ with $N \ge 2$).

	The two-photon JC Hamiltonian could be implemented via dispersive dipolar interaction involving a three-level superconducting artificial atom, similar to the scheme performed in Ref.~\cite{Wang2016}, where an effective coupling of the order of $1$ MHz ($\sim 10^3 \kappa$) was achieved, which would be sufficient to observe single-photon filter and rotations involving $\ket{0}$ and $\ket{1}$, as seen in Fig.~\ref{Figure_rotations}(a) [a complete rotation is achieved in a time ($\sim 30$ $\mu$s) 2 orders of magnitude shorter than the single-photon cavity lifetime, $1/\kappa \sim 1$ ms]. 
	


	\emph{Weak coupling regime: Multiphoton-induced-reflectivity phenomenon.}---A weak driving field, of constant strength $\varepsilon(t) = \varepsilon_{0} \ll g$, frequency $\omega_{p}$, and the same nonlinearity of the JC Hamiltonian ($M=N)$, will be able to introduce $N$ excitations to the system if it is close to resonance with the transitions $\ket{g,0} \leftrightarrow \left|\pm,N\right\rangle $, i.e., if $N\omega_{p}\approx\omega N\pm g\sqrt{N!}$ (for $\omega_{0}=N\omega$). Nonetheless, if the decay rates of the first dressed states are large enough when compared to the effective Rabi frequency $g\sqrt{N!}$, an interference effect due to the different absorption paths can take place, canceling out the system absorption. This phenomenon is analogous to the electromagnetically induced transparency (EIT) that happens in three-level atoms driven by probe and control fields \cite{Fleischhauer2005}. In our case, however, the system does not become transparent to the probe field, it becomes highly reflective instead, as briefly discussed by the authors for $M=N=1$ in Ref.~\cite{Rossatto2013}, which gives rise to a multiphoton-induced-reflectivity phenomenon.
	
\begin{figure}[b]
		\includegraphics[trim = 6mm 1mm 1mm 1mm, clip, width=0.48\textwidth]{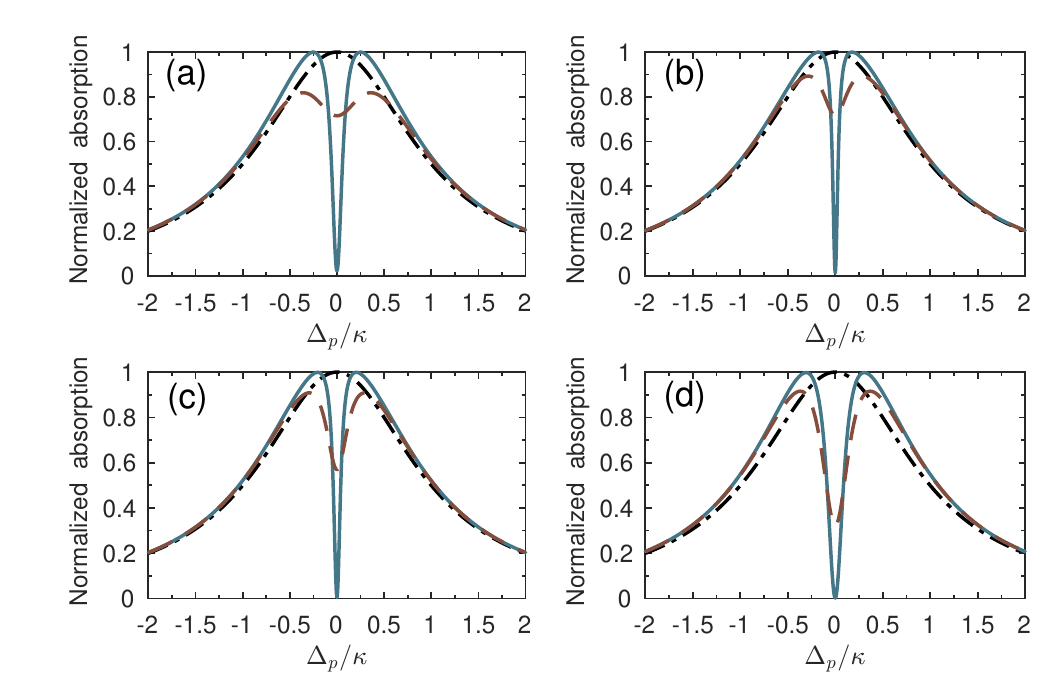}
		\caption{Normalized absorption spectrum of the cavity mode. The black dash-dotted lines are for $g=0$, while $g=\kappa/4$ with $\gamma=\gamma_\phi=10^{-4}\kappa$ (blue full lines) and $\gamma=\gamma_\phi=\kappa/10$ (red dashed lines). We have adjusted the strength of the continuous probe field $\varepsilon_0 = g/10\sqrt{N!}$ to keep the maximum average number of photons of the order of $10^{-3}$ for all cases: (a) nonlinearity $N=M=1$, (b) $2$, (c) $3$, and (d) $4$.} 
		\label{Figure_EIT}
	\end{figure}

	To illustrate this effect, in Fig.~\ref{Figure_EIT} we show the normalized mean number of intracavity excitations, i.e., the normalized absorption ($\langle a^{\dagger}a\rangle/\langle a^{\dagger}a\rangle_{\rm max}$), as a function of the detuning between the probe and cavity-mode frequencies $\Delta_{p}$ for different nonlinearities, with $\moy{a^{\dagger}a}_{\rm max}$ calculated with $g=\Delta_p=0$. All curves were obtained by taking the steady-state solution of the master equation [Eq.~\eqref{eq:masterequation}].

	We notice that the stronger the atomic dissipation rates, the less the cancellation of the absorption of the probe field on the resonance ($\Delta_{p}=0$). This happens because the cavity decay rate ($\kappa$) plays the same role as the excited-state decay rate in usual EIT experiments, while both our $ \gamma$ and $\gamma_\phi$ play the same role as the dephasing rate \cite{Souza2015}, which is an agent that obstructs the destructive interference responsible for the cancellation of the absorption of the probe field in the transparency window (reflectivity window in our case). We can also observe that the width of the reflectivity window increases when we increase $N$. This is due to fact that the width depends not only on $g$, but rather on the effective vacuum Rabi frequencies $g \sqrt{N!}$, similar to the case in which there are many atoms in cavity-EIT experiments \cite{Mcke2010}. Thus, here we identify that the effective atom-mode coupling in our case corresponds to the Rabi frequency of the control field in usual EIT experiments \cite{Souza2015}. It is worth noting that in our case the destructive interference is not due to an interference between two single-photon absorption paths as happens in usual EIT phenomenon \cite{Fleischhauer2005} and analogues \cite{Rossatto2013,Souza2015}, but due to an interference between two cascade decay channels.
	
	Finally, a multiphoton JC system in this configuration can be seen as a quantum-optical frequency blocker, i.e., a quantum device (filter) that inhibits the passage of multiphoton external probe fields within a narrow (controllable) reflectivity window in frequency domain. In addition, our study paves the way to further investigation (and possible applications) of processes involving more complex multiphoton quantum interference, e.g., the extension of the cavity-EIT phenomenon, which considers three-level atoms, to the case in which the atom-cavity interaction is nonlinear.

	\emph{Conclusion.}---We have investigated a multiphoton Jaynes-Cummings system driven by a nonlinear driving field. Working in the strong atom-cavity coupling regime, we have shown that the cavity dynamics can be restricted to an upper-limited Fock subspace with the atom kept in its ground state. Thus, by driving the cavity mode with a coherent field (superposition of many Fock states), only the cavity states within the aforementioned Fock subspace can be populated; i.e., we have shown how to implement a quantum scissor without requiring a nonlinear Kerr medium inside the cavity. From the point of view of the cavity transmission, this Fock state filter can also be seen as a quantum-optical intensity limiter. Additionally, the nonlinearities of the system and driving field can be chosen in order to allow arbitrary rotations in the Fock space of a bosonic mode. On the other hand, working in the weak atom-cavity coupling regime, we have shown that the system exhibits a multiphoton-induced-reflectivity phenomenon, i.e., the multiphoton Jaynes-Cummings interaction induces a narrow (controllable) reflectivity window in frequency domain for multiphoton driving fields, such that this system can be used as a quantum-optical frequency blocker. Our results are useful for applications in quantum information protocols that require arbitrary rotations in Fock space \cite{Santos2005}, the generation of finite-dimensional quantum-optical states \cite{Leoski2001} and the manipulation of the optical response of atom-field systems \cite{Souza2013b}. As discussed, our results are general and could be implemented at least for two-photon interactions and driving fields with current technology \cite{Vogel1995,Meekhof1996,Gau1992,brune1997,neil2015,Garziano2015,Bertet2005,Felicetti2018,Feli2018}. However, the experimental verification for higher nonlinearities remains open, which should motivate further research on this topic, also motivated by some possible applications, e.g., the use of an $N$-photon Jaynes-Cummings system as a qudit with harmonic spectrum, a deeper investigation of the multiphoton blockade in the system or even how to use it as a single-photon source, and an exploration of multiphoton quantum interference for more-complex processes.

	\begin{acknowledgments}
		We thank M. H. Oliveira for the cavity drawing. This work was supported by the S\~{a}o Paulo Research Foundation (FAPESP) Grants No.~2013/04162-5 and No.~2013/23512-7, National Council for Scientific and Technological Development (CNPq) Grant No.~308860/2015-2, and Brazilian National Institute of Science and Technology for Quantum Information (INCT-IQ) Grant No.~465469/2014-0.
	\end{acknowledgments}

	\bibliography{bibliography}



\section*{Supplementary material}

# Supplemental Material for Multiphoton Jaynes-Cummings Model: Arbitrary Rotations in Fock Space and Quantum Filters

## I. DERIVATION OF THE EFFECTIVE HAMILTONIAN

Consider the multiphoton Jaynes-Cummings Hamiltonian ($\hbar = 1$)

$$H_0 = \omega a^\dagger a + \omega_0 \frac{\sigma_z}{2} + g(\sigma_+ a^N + \text{H.c}), \tag{1}$$

with $\omega$ and $a$ ($a^\dagger$) being the frequency and the annihilation (creation) operator of the single-mode bosonic field. The atomic frequency transition between the ground $|g\rangle$ and the excited $|e\rangle$ states is $\omega_0$, while $\sigma_+ = (\sigma_-)^\dagger = |e\rangle\langle g|$ and $\sigma_z = \sigma_+\sigma_- - \sigma_-\sigma_+$ are atomic operators (spin-$\frac{1}{2}$ Pauli matrices). The vacuum Rabi frequency (atom-field coupling) is $g$, the nonlinearity of the JC model is the positive integer $N$, and H.c. stands for Hermitian conjugate.

The lowest eigenstates of the system are uncorrelated, involving the atomic ground state and up to $N-1$ excitations in the bosonic mode, namely (with the corresponding eigenenergies)

$$|\Psi_{g,n}\rangle = |g, n\rangle, \ E_{g,n} = \left(n - \frac{N}{2}\right)\omega, \tag{2}$$

with $0 \leq n < N$. The remaining eigenstates or dressed states (eigenenergies), which are correlated, are

$$|\pm, n\rangle = \frac{1}{\sqrt{2}}\left(|g, n\rangle \pm |e, n - N\rangle\right), \tag{3}$$

$$E_{\pm,n} = \left(n - \frac{N}{2}\right)\omega \pm g\sqrt{\frac{n!}{(n-N)!}}, \tag{4}$$

with $n \geq N$. The closure relation of the eigenstates of $H_0$ is

$$I = \sum_{n=0}^{N-1} |g, n\rangle\langle g, n| + \sum_{n=N}^{\infty} \left(|+, n\rangle\langle +, n| + |-, n\rangle\langle -, n|\right). \tag{5}$$

Consider that the system is subject to a driving field of strength $\varepsilon(t)$, single-photon frequency $\omega_p$ and phase $\chi$, which is described by $H_P = \varepsilon(t)\left[a^M e^{i(M\omega_p t - \chi)} + \text{H.c.}\right]$, with $M$ being the nonlinearity of the driving field. Thus, the total Hamiltonian is $H = H_0 + H_P$. To understand what can happen with our system, it is convenient to derive an effective Hamiltonian in the limit of strong atom-field coupling when compared with the probe field strengh, i.e., when $g \gg \max|\varepsilon(t)|$. To this end let us rewrite $H_P$ (for $M < N$) with the help of the closure relation given by Eq. (5),

$$H_p = \varepsilon(t)\left[a^M e^{i(M\omega_p t - \chi)} + \text{H.c.}\right]\left\{\sum_{n=0}^{N-1} |g, n\rangle\langle g, n| + \sum_{n=N}^{\infty}\left(|+, n\rangle\langle +, n| + |-, n\rangle\langle -, n|\right)\right\}$$

$$= \varepsilon(t)e^{i(M\omega_p t - \chi)}\left\{\sum_{n=M}^{N-1} a^M |g, n\rangle\langle g, n| + \sum_{n=N}^{\infty} a^M \left(|+, n\rangle\langle +, n| + |-, n\rangle\langle -, n|\right)\right\} + \text{H.c.}$$

$$= \varepsilon(t)e^{i(M\omega_p t - \chi)}\left\{\sum_{n=M}^{N-1} a^M |g, n\rangle\langle g, n| + \sum_{n=N}^{\infty} a^M \left(|g, n\rangle\langle g, n| + |e, n - N\rangle\langle e, n - N|\right)\right\} + \text{H.c.}$$

$$= \varepsilon(t)e^{i(M\omega_p t - \chi)}\left\{\sum_{n=M}^{N-1} \sqrt{\frac{n!}{(n-M)!}}|g, n - M\rangle\langle g, n| + \underbrace{\sum_{n=N}^{\infty} \sqrt{\frac{n!}{(n-M)!}}|g, n - M\rangle\langle g, n|}\right.$$

$$+ \sum_{n=N+M}^{\infty} \sqrt{\frac{(n-N)!}{(n-N-M)!}} |e, n-N-M\rangle \langle e, n-N| \Bigg\} + \text{H.c.}$$

$$= \varepsilon(t) e^{i(M\omega_p t - \chi)} \left\{ \sum_{n=M}^{N-1} \sqrt{\frac{n!}{(n-M)!}} |g, n-M\rangle \langle g, n| + \sum_{n=N}^{N+M-1} \sqrt{\frac{n!}{(n-M)!}} |g, n-M\rangle \langle g, n| \right.$$

$$\left. + \sum_{n=N+M}^{\infty} \sqrt{\frac{n!}{(n-M)!}} |g, n-M\rangle \langle g, n| + \sum_{n=N+M}^{\infty} \sqrt{\frac{(n-N)!}{(n-N-M)!}} |e, n-M-N\rangle \langle e, n-N| \right\} + \text{H.c.}.$$

The underlined term was split in 2 sums because one part is in the subspace of the decoupled states and the other part is in the subspace of the coupled states. Rewriting the states in the Hamiltonian above in terms of the eigenstates of the Hamiltonian $H_0$ give us

$$H_p = \varepsilon(t) e^{i(M\omega_p t - \chi)} \left\{ \sum_{n=M}^{N-1} \sqrt{\frac{n!}{(n-M)!}} |g, n-M\rangle \langle g, n| + \sum_{n=N}^{N+M-1} \sqrt{\frac{n!}{(n-M)!}} |g, n-M\rangle [\langle +, n| + \langle -, n|] \right.$$

$$+ \frac{1}{2} \sum_{n=N+M}^{\infty} \sqrt{\frac{n!}{(n-M)!}} [|+, n-M\rangle + |-, n-M\rangle] [\langle +, n| + \langle -, n|]$$

$$\left. + \frac{1}{2} \sum_{n=N+M}^{\infty} \sqrt{\frac{(n-N)!}{(n-N-M)!}} [|+, n-M\rangle - |-, n-M\rangle] [\langle +, n| - \langle -, n|] \right\} + \text{H.c.}.$$

Now we are ready to analyze the effective dynamics of the system under the action of a driving field. Through the unitary transformation

$$U_0 = \exp(-iH_0 t),$$

the total Hamiltonian of the system can be transformed as

$$H_T = U_0^{-1} H U_0 - H_0 = U_0^{-1} H_P U_0$$

$$= \varepsilon(t) e^{i(M\omega_p t - \chi)} \left\{ \sum_{n=M}^{N-1} \sqrt{\frac{n!}{(n-M)!}} U_0^{-1} |g, n-M\rangle \langle g, n| U_0 \right.$$

$$+ \frac{1}{\sqrt{2}} U_0^{-1} \sum_{n=N}^{N+M-1} \sqrt{\frac{n!}{(n-M)!}} |g, n-M\rangle [\langle +, n| + \langle -, n|] U_0$$

$$+ \frac{1}{2} \sum_{n=N}^{\infty} U_0^{-1} \sqrt{\frac{n!}{(n-M)!}} [|+, n-M\rangle + |-, n-M\rangle] [\langle +, n| + \langle -, n|] U_0$$

$$\left. + \frac{1}{2} \sum_{n=N+M}^{\infty} U_0^{-1} \sqrt{\frac{(n-N)!}{(n-N-M)!}} [|+, n-M\rangle - |-, n-M\rangle] [\langle +, n| - \langle -, n|] U_0 \right\} + \text{H.c.}.$$

Then

$$H_T = \varepsilon(t)e^{i(M\omega_p t - \chi)}\left\{\sum_{n=M}^{N-1}\sqrt{\frac{n!}{(n-M)!}}|g, n-M\rangle\langle g, n|e^{iE_{g,n-M}t - iE_{g,n}t}\right.$$

$$+\frac{1}{\sqrt{2}}\sum_{n=N}^{N+M-1}\sqrt{\frac{n!}{(n-M)!}}\left[|g, n-M\rangle\langle +, n|e^{iE_{g,n-M}t - iE_{+,n}t} + |g, n-M\rangle\langle -, n|e^{iE_{g,n-M}t - iE_{-,n}t}\right]$$

$$+\frac{1}{2}\sum_{n=N}^{\infty}\sqrt{\frac{n!}{(n-M)!}}\left[|+, n-M\rangle\langle +, n|e^{iE_{+,n-M}t - iE_{+,n}t} + |-, n-M\rangle\langle +, n|e^{iE_{-,n-M}t - iE_{+,n}t}\right.$$

$$\left. + |+, n-M\rangle\langle -, n|e^{iE_{+,n-M}t - iE_{-,n}t} + |-, n-M\rangle\langle -, n|e^{iE_{-,n-M}t - iE_{-,n}t}\right]$$

$$+\frac{1}{2}\sum_{n=N+M}^{\infty}\sqrt{\frac{(n-N)!}{(n-N-M)!}}\left[|+, n-M\rangle\langle +, n|e^{iE_{+,n-M}t - iE_{+,n}t} - |-, n-M\rangle\langle +, n|e^{iE_{-,n-M}t - iE_{+,n}t}\right.$$

$$\left.\left. - |+, n-M\rangle\langle -, n|e^{iE_{+,n-M}t - iE_{-,n}t} + |-, n-M\rangle\langle -, n|e^{iE_{-,n-M}t - iE_{-,n}t}\right]\right\} + \text{H.c.}.$$

Now, adjusting $\omega_P = \omega$ and substituting the eigenenergies given above, the Hamiltonian becomes

$$H_T = \varepsilon(t)e^{-i\chi}\left\{\sum_{n=M}^{N-1}\sqrt{\frac{n!}{(n-M)!}}|g, n-M\rangle\langle g, n|\right.$$

$$+\frac{1}{\sqrt{2}}\sum_{n=N}^{N+M-1}\sqrt{\frac{n!}{(n-M)!}}\left[|g, n-M\rangle\langle +, n|e^{-igt\sqrt{\frac{n!}{(n-N)!}}} + |g, n-M\rangle\langle -, n|e^{igt\sqrt{\frac{n!}{(n-N)!}}}\right]$$

$$+\frac{1}{2}\sum_{n=N+M}^{\infty}\sqrt{\frac{n!}{(n-M)!}}\left[|+, n-M\rangle\langle +, n|e^{igt\left(\sqrt{\frac{(n-M)!}{(n-M-N)!}} - \sqrt{\frac{n!}{(n-N)!}}\right)} + |-, n-M\rangle\langle +, n|e^{-igt\left(\sqrt{\frac{(n-M)!}{(n-M-N)!}} + \sqrt{\frac{n!}{(n-N)!}}\right)}\right.$$

$$\left. + |+, n-M\rangle\langle -, n|e^{igt\left(\sqrt{\frac{(n-M)!}{(n-M-N)!}} + \sqrt{\frac{n!}{(n-N)!}}\right)} + |-, n-M\rangle\langle -, n|e^{-igt\left(\sqrt{\frac{(n-M)!}{(n-M-N)!}} - \sqrt{\frac{n!}{(n-N)!}}\right)}\right]$$

$$+\frac{1}{2}\sum_{n=N+M}^{\infty}\sqrt{\frac{(n-N)!}{(n-N-M)!}}\left[|+, n-M\rangle\langle +, n|e^{igt\left(\sqrt{\frac{(n-M)!}{(n-M-N)!}} - \sqrt{\frac{n!}{(n-N)!}}\right)} - |-, n-M\rangle\langle +, n|e^{-igt\left(\sqrt{\frac{(n-M)!}{(n-M-N)!}} + \sqrt{\frac{n!}{(n-N)!}}\right)}\right.$$

$$\left.\left. - |+, n-M\rangle\langle -, n|e^{igt\left(\sqrt{\frac{(n-M)!}{(n-M-N)!}} + \sqrt{\frac{n!}{(n-N)!}}\right)} + |-, n-M\rangle\langle -, n|e^{-igt\left(\sqrt{\frac{(n-M)!}{(n-M-N)!}} - \sqrt{\frac{n!}{(n-N)!}}\right)}\right]\right\} + \text{H.c.},$$

which contains highly oscillating terms. Now, performing an RWA we end up with

$$H_{\text{eff}} \simeq \varepsilon(t)e^{-i\chi}\sum_{n=M}^{N-1}\sqrt{\frac{n!}{(n-M)!}}|g, n-M\rangle\langle g, n| + \text{H.c.}.$$

Since we are interested in the cases in which the system is initially in its ground state, the validity of this approximate Hamiltonian can be reduced to the condition

$$\max|\varepsilon(t)| \ll g\sqrt{2(N-M)!},$$

which is the condition for neglecting (via the RWA) the term related to the transitions $|g,0\rangle \to |\pm,1\rangle$. Being this the case, the doublets will practically not be populated and therefore the conditions to neglect (via the RWA) the remaining terms related to the transitions between the doublets become irrelevant, since such transitions will have extremely small probabilities of occurring.

Examples:

i) $N=2$ and $M=1$: $\max|\varepsilon(t)| \ll g\sqrt{2}$. In this case, the transformed Hamiltonian will be

$$H_T \simeq \varepsilon(t)e^{-i\chi}\sum_{n=1}^{1}\sqrt{\frac{n!}{(n-M)!}}|g,n-M\rangle\langle g,n| + \text{H.c.} = \varepsilon(t)e^{-i\chi}|g,0\rangle\langle g,1| + \text{H.c.},$$

which describes a perfect rotation involving the Fock states $|0\rangle$ and $|1\rangle$.

ii) $N=3$ and $M=2$: $\max|\varepsilon(t)| \ll g\sqrt{2}$. In this case, the transformed Hamiltonian will be

$$H_T \simeq \varepsilon(t)e^{-i\chi}\sum_{n=2}^{2}\sqrt{\frac{n!}{(n-M)!}}|g,n-M\rangle\langle g,n| + \text{H.c.} = \varepsilon(t)e^{-i\chi}\sqrt{2}|g,0\rangle\langle g,2| + \text{H.c.},$$

which describes a perfect rotation involving the Fock states $|0\rangle$ and $|2\rangle$.

iii) $N=3$ and $M=1$: $\max|\varepsilon(t)| \ll 2g$. In this case, the transformed Hamiltonian will be

$$H_T \simeq \varepsilon(t)e^{-i\chi}\sum_{n=1}^{2}\sqrt{\frac{n!}{(n-M)!}}|g,n-M\rangle\langle g,n| + \text{H.c.} = \varepsilon(t)e^{-i\chi}|g,0\rangle\langle g,1| + \varepsilon(t)e^{-i\chi}\sqrt{2}|g,1\rangle\langle g,2| + \text{H.c.}.$$

This Hamiltonian indicates that the cavity mode can only be populated in 0, 1, and 2 photon states.



\end{document}